# A Holistic Study of the W UMa Binary EQ Tau

M. M. Elkhateeb[1,2] and M. I. Nouh[1,2]

[1]Astronomy Department, National Research Institute of Astronomy and Geophysics, 11421 Helwan, Cairo, Egypt

E-mail: abdo_nouh@hotmail.com

Fax: +202 2554 8020

[2]Physics department, College of Science, Northern Border University, 1321 Arar, Saudi Arabia

**Abstract**: We present a new BVR light curves of the system EQ Tau carried out in the period from March to April 2006 using a 50-cm F/8.4 Ritchey–Chretien telescope (Ba50) of the Baja Astronomical Observatory (Hungary), and 512 × 512 Apogee AP-7 CCD camera. The observed light curves were analyzed using the 2009 version of the Wilson-Devinney code. The results show that the more massive component is hotter than the low massive one by 99 $^0$K. A long term orbital period study show that the period increases by the rate 8.946 x $10^{-11}$ day/cycle. Evolutionary state of the system has been investigated and showed that, the primary component of the system is located nearly on the ZAMS for both the M-L and M-R relations. The secondary component is close to the TAMS track for M-L and above the M-R relations.

## 1. Introduction

The variability of the system EQ Tau (G2, V=12.04, P=0.341349) was discovered by Tesevich (1954). The system was included to the AAVSO list of eclipsing binaries (Baldwin & Samolyk 1993). The system was neglected for over two decades. The first period was calculated by Whitney (1972), while the first CCD light curve was carried out in R band by Benbow and Mutel (1995). The system was monitored by Buckner, Nellermoe & Mutel (1998), and Nelson (2001).

The first reliable spectroscopic elements of the system were estimated by Rucinski et al. (2001). Successive observations were carried out from 2000 to 2002 in BV band pass by Pribulla and Vanko (2002), and Vanko et al. (2004). They combined the calculated photometric elements resulting from their observations with published spectroscopic elements to yield the absolute parameters of the system. Light curves asymmetry was noted by Yang and Liu (2002) through their BV observations and they adopted the first spot model for the system EQ Tau. Many photometric studies were applied for the system light curve by Hrivnak et al. (2006), Yuan and Qian (2007) and Alton (2006, 2009). Recently, during the work in the present analysis, Li et al. (2014) presents a new light curve analysis for the system.

In the present paper we are going to use the Wilson-Devinney (WD) code to estimate the physical parameters of the contact binary EQ Tau and to study its evolutionary status.

## 2. Observations

Observations of EQ Tau were carried out on five nights from March to April 2006 in V and R bandpass using a 50-cm F/8.4 Ritchey–Chretien telescope (Ba50) of the Baja Astronomical Observatory (Hungary), and $512 \times 512$ Apogee AP-7 CCD camera. Table (1) listed the coordinates of the variable, the comparison and the check star. It's clear that the comparison and check stars are close to the variable and they were in the same field which leads to ignore the extinction corrections. The observed images were analyzed by the photometry software AIP4WIN (Berry and Buruell 2000) which based on aperture photometry, including bias and dark subtraction and flat field correction. A total of 649 individual observations were obtained in VR band pass (252 in V and 397 in R), which covered a complete light curves in V and R band, and displayed in Figure (1). The individual observations were listed in Table 2, as heliocentric Julian dates and phases together with the magnitude difference (the variable minus the comparison). The phases were calculated using the ephemeris of Kreiner et al. (2000):

$$\text{Min I} = 2440203.4343 + 0.^{d}341347947 * E \qquad (1)$$

From the present observation, the times of three primary light minima were calculated using the Minima V2.3 Package (Nelson 2006) based on the Kwee & Van Worden (1956) fitting method. The calculated minima appear in Table (3) and together in Table (4) with all published photoelectric and CCD minima.

## 3. Orbital Period Behavior

As mentioned in the last section, the system EQ Tau was neglected since its discovery. Qian and Ma (2001) showed that the period of the system EQ Tau decreases by the rate -1.72 x $10^{-7}$ day/year for 23 year before. Yang and Liu (2002), showed that the orbital period of the system exhibit a wavelike variation with cycle variation of 23 yr they refer the periodic change to a strong magnetic field or the presence of a third body orbits the system EQ Tau. Pribulla and Vanko (2002) interpreted the long term period changes by the presence of a third body on a 50 yr orbit. Alton (2006) estimated a very slow orbital period increase with a rate of 0.021 sec/year. Hrivnak et al. (2006) studied the period variation using the list of minima collected by Pribulla and Vanko (2002). They showed that the periodicity adopted by Pribulla and Vanko (2002) was based on older visual data of Tsesevich (1954) from 1954 and photographic minima of Whitney (1972) from 1959 to 1962, while the period behavior based on photoelectric and recent photographic data showed a constant trend. They re-analyzed the period stability using high quality timings of minima of photoelectric observations together with their observed minima covering the interval from 1989 to 2005. Their results showed that the period of the system EQ Tau changes from one constant period to another around 1972 instead of cyclical variation.

A cyclical behavior for the orbital period of the system EQ Tau was announced again by Yuan and Qian (2007) with a periodic variation of period 48.5 yr based on all published minima covering 64 yr. Alton (2009) showed that the period increased by 0.015 sec/yr over the interval from 2000-2009. From the previous orbital period studies of the system EQ Tau, it's clear that the trend of period behavior depend on the quality of the collected minima and the covered interval. In the present paper we used an updated list of published minima from the literatures together with that listed in the web site (http:J//astro.sci.muni.cz/variables/ocgate/) and our new observed minima, which produced a complete set of data. A total of 264 timings of minima covered the interval of 59 yr (~ 7351 revolutions) from 1954 to 2013 were used to follow the long term behavior of the system EQ Tau by means of (O-C) diagram.

Linear ephemeris of Kreiner et al. (2000) (Eq. 1) was used to determine the calculated "C" values of the eclipse timings, listed in Table (4). Some collected uncertain minima were discarded in order to estimate accurate results. The (O-C) values were represented in Figure (2) versus the integer cycle E, where no distinctions have been made between primary and secondary minima. Scattering of some minima in the (O-C) diagram may resulted from the variation in the observed light curves, which leads to non-asymmetry and also uncertainty in the calculated minima. The trend of the (O-C) diagram shows that the behavior of the orbital period of the system EQ Tau changes between increasing and decreasing since its discovery and can't be represented by any mean elements derive by the linear best fit. In order to describe and follow the period behavior of the system EQ Tau through the last 71 years, since its discovery we divided the (O-C) diagram into four intervals $E_i - E_{i-1}$, i = 1, 2, 3, 4. The four intervals and their best fit period results data (with corresponding change in the period $\Delta p$) were listed in Table (5) together with the standard deviations **SD**, correction coefficient **r** and the residual some of squares. Results in Table (5) reveal that the period of the system EQ Tau shows two intervals of increase and similar of decrease, which looks like as a periodic behavior. The general trend of the (O-C) diagram can be represented by a fifth degree polynomial with residual sum of square = 0.0014 and correlation coefficient = 0.918 as:

$$\text{Min I} = 2440203.4314 + 0.341349197 * E - 4.4731*10^{-11} * E^2 - 1.7186*10^{-15} E^3$$

$$+ 6.4851*10^{-20} * E^4 - 4.4607*10^{-25} * E^5 \qquad (2)$$

Eq. 2 represented a new light element for the system EQ Tau which shows a period increase with rate $dP/dE = 8.946 \times 10^{-11}$ day/cycle or $9.566 \times 10^{-8}$ day/year or 0.83 second/century. The (O-C)p residuals were calculating using polynomial ephemeris (Eq. 2) and listed in Table (4) and displayed in Figure (2).

## 4. Light Curve Variation

The historical survey of the published light curves for the system EQ Tau since its discovery showed asymmetry and distortions at maximum phases, which seem to be referred to surface in-homogeneities of the components. The primary maximum (Max I) is brighter than the secondary one (Max II), which known as O'Connell effect and appears in many eclipsing binaries and may refer to hot or cool regions on either components. Observations by Yang and Liu (2002) showed a typical O'Connell effect and a primary maximum brighter than secondary one by 0.03 mag., while Pribulla and Vanko (2002) showed maximum difference of about 0.017. Observed light curves by Hrivnak et al (2006) displayed a flat-bottom in the secondary minimum which indicates that the cooler component was eclipsed totally and showed a maximum differences (O'Connell effect) of about ~ 0.03 mag. Alton (2006, 2009) observed an apparent asymmetry at Max I and refers it to hot or dark spot on either components facing the observer.

A noticeable variation in the three light curves was observed by Yuan and Qian (2007). They explained this variation to a strong and variable dark spot activity. Most of the previous observations together with our light curves indicates some distortions in the two maximum phase (Max I and Max II) and also magnitude difference between them (O'Connell effect). This phenomena was observed in many contact binary (i.e. AQ Tuc (Hilditch, King (1986), and CN And Keskin (1989)), they explained it as a pulsation of a common envelope due to mass transfer between two components (Li et al. 2002). The magnitude difference between the two maxima and apparent asymmetry may be arisen from hot star spot(s) on either components (which increases the flux during Max I) Alton (2006), or dark star spot(s) which decrease the level of Max II. A relation between the light curve variation and orbital period changes in the same cycle was predicted by Appliegate (1992).

Using our observations together with all published light curves, the light levels (Max I, Max II, Min I, and Min II) were estimated. The amplitude (depths) of the primary $A_p$ (mag(Min I – Max I)) and secondary $A_s$ (mag(Min II – Max I)) and the magnitude difference between both maxima (O'Connell effect) $D_{max}$ (mag(Max I – Max II)) and minima $D_{min}$ (mag(Min I – Min II)) have been calculated for each light curve and listed in Table (6), while displayed in Figure (4). The behavior of the parameters $D_{max}, D_{min}, A_p,$ and $A_s$ showed a wave-like variation as period function. This behavior may be interpreted by a periodic effect of some physical mechanism. It's clear that both the period change behavior and light curve parameters ($D_{max}, D_{min}, A_p,$ and $A_s$) of the system EQ Tau showed a periodic variation which may be referred to the presence of magnetic activity cycles in more massive components and/or mass transfer mechanism. A future sequence series of observations for the system EQ Tau are needed in order to follow this periodic variation. Variation of the brightness difference between both maxima ($D_{max}$) and minima ($D_{min}$) (Fig. 6. a,b.) can be represented by a fitting as seen in Eq. 3 and 4 respectively as:

$$D_{max} = -0.0210 + 0.0356 * COS(2.2417*t - 150.7340) \qquad (3)$$
$$D_{min} = 0.0525 + 0.2886 * COS(0.5131*t - 2.9724) \qquad (4)$$

Behavior of the amplitude (depth) of the primary $A_p$ and secondary $A_s$ (Fig. 6. c, d) can be represented by a sinusoidal fit as seen in Eq. 5 and 6 respectively as:

$$A_p = 0.6230 + 0.1723 * COS(0.6870*t - 345.4671) \qquad (5)$$

$$A_s = 0.5617 + 0.1249 * COS(0.6905*t - 352.2458) \qquad (6)$$

## 5. Light Curve Modeling

Although the system EQ Tau was discovered since 1954 (60 years ago) it didn't take sufficient interest in observation and light curve modeling. The first photometric solution was published by Pribulla and Vanko (2002) based on photoelectric observations in BV band pass. Although the asymmetry was clear in the observed light curves, they ruled out any spotted solution due to the low quality of their observations. A series of photometric solution were announced by Yang & Liu (2002), Vanko et al. (2004), Hrivnak et al. (2006), Alton (2006 and 2009), Yuan and Qian (2007) and I t al. (2014).

Because of the clear asymmetry in all published light curves of the system EQ Tau the previous photometric solutions are in agree on the presence of spot(s) on either or both system components. The only difference was the type of the spot(s) (hot or cool) and their distribution on the star's surface.

In this paper we present a photometric solution using WDint56a Package (Nelson 2009) which is a windows interface synthetic light curve and differential correction package of 2009 version of Wilson and Devinney (W-D) code. We analyzed the individual light curve observations instead of normal light curves, which don't reveal a real light variation of the system. The surface temperature of the primary star was fixed at 5800 $^0K$ and the logarithm law of Van Hamme (1993) table was used to adopt and interpolate the bolometric limb darkening coefficients ($x_1 = x_2$, $y_1 = y_2$) and model atmosphere was applied. We adopted $g_1 = g_2 = 0.32$ (Lucy 1967) and the albedo value $A_1 = A_2 = 0.5$ (Rucinski 1969). During the execution of WDint56a Package, Mode 3 (overcontact) was applied and some parameters were held fixed (i.e. $T_1$, g, A, $x_1 = x_2$). The adjusted parameters are the temperature $T_2$ of the secondary component, the surface potentials $\Omega_1 = \Omega_2$, the mass ration q ($M_2/M_1$), and the monochromatic luminosity $L_1$ of star 1 (the relative luminosity of star 2 was calculated by the stellar atmosphere model). The accepted model reveals modified parameters which listed in Table (7) and represented in Figure (5). It's clear that the accepted model included three hot spots (two on the primary component and only one on the secondary, also the more massive component is hotter than the low massive one by $\Delta T \sim 99\ ^0K$. Based on the calculated parameters a three dimension geometric structure of the system EQ Tau was constructed using the software package Binary Maker 3.03 (Bradstreet and Steelman 2004) (Figure (6)). The absolute physical dimensions were calculated and listed in Table 8 together with all published parameters calculated by previous photometric solutions.

## 6. Discussion and Conclusion

New VR CCD observations were carried out for the system EQ Tau which added a new five minima. Using all published times of minima together with our new minima to study a long term orbital period behavior of the system EQ Tau, which showed a periodical variation with a periodicity of about 43.3 yr. The period increase with a rate of $dP/dE = 8.946 \times 10^{-11}$ day/cycle or $9.566 \times 10^{-8}$ day/year or 0.83 second/century. Long term stability of the light curves of the system EQ Tau shows a periodic change in the magnitude difference between both maxima $D_{max}(t)$ (O'Connell effect) and minima $D_{min}(t)$, the amplitude (depths) of the primary $A_p(t)$ and secondary $A_s(t)$. A synchronous variation in both orbital period and light curve parameters $D_{max}(t)$, $D_{min}(t)$, $A_p(t)$ and $A_s(t)$ for the system EQ Tau may be interpreted by the presence of magnetic activity cycle and/or mass transfer mechanism. A photometric solution of the observed light curves by means of W-D code showed that the more massive component is hotter than the low massive one by $\sim 99\ ^0K$.

We used the physical parameters listed in Table (8) to investigate the current evolutionary status of EQ Tau. In Figures (7) and (8), we plotted the components of EQ Tau on the mass–luminosity (M-L) and mass–radius (M-R) relations along with the evolutionary tracks computed by Girardi et al. (2000) for both zero age main sequence stars (ZAMS) and terminal age main sequence stars (TAMS) with metalicity $z = 0.019$. As it is clear from the figures, the primary component of the system is located nearly on the ZAMS for both the M-L and M-R relations. The secondary component is close to the TAMS track for M-L and above the M-R relations. For the sake of comparison, we plotted sample of W-type contact binaries listed in Table (9). The components of EQ Tau have the same behavior of the selected W-type systems.

To locate components on the $T_{eff}$-luminosity relation for singles stars, we used for this purpose, the non-rotated evolutionary models of Ekström et al. (2012) in the range 0.8–120 $M_\odot$ at solar metalicity (z = 0.014). W used the tracks for the masses 0.9 $M_\odot$, 1 $M_\odot$ and 1.1 $M_\odot$. The highly poor fit of the secondary and the fair fit of the primary reflect the contact nature of the system.

The mass-effective temperature relation (M–$T_{eff}$) relation for intermediate and low mass stars (Malkov, 2007) is displayed in Figure (10 ). The location of our mass and radius on the diagram revealed a good fit for the primary and poor fit for the secondary components. This gave the same behavior of the system on the mass-luminosity and mass-radius relations.

Because of that EQ Tau is of W-type, its evolutionary state has great interest. According to the scenario proposed by Qian (2001), EQ Tau with mass ratio q=0.445 suffers from mass transfer from the secondary to the primary. After that it oscillates around q=0.4 and reverse the mass transfer from the primary to the secondary. To follow this oscillation and the evolution of the system, one needs continuous observations of the system.

Table 1. Coordinates of EQ Tau, comparison, and the check stars

| Star Name | (2000.0) | δ (2000.0) | V | B-V |
|---|---|---|---|---|
| EQ Tau | 03$^h$ 48' 16.0" | +22$^o$ 17' 30.0" | 11.18 | 0.86 |
| Comparison (TYC 1260-575-1) | 03$^h$ 48' 16.5" | +22$^o$ 17' 29.7" | 10.32 | 1.15 |
| Check (GSC 01260-00800) | 03$^h$ 48' 17.4" | +22$^o$ 22' 43.0" | 8.190 | 1.05 |

Table 2. CCD Observations of EQ Tau in VR band.

| V-band observations of EQ Tau | | | | R-band observations of EQ Tau | | | |
|---|---|---|---|---|---|---|---|
| JD | Phase | ΔV | Error | JD | Phase | ΔR | Error |
| 2453796.3086 | 0.0821 | 0.962 | 0.0393 | 2453796.3077 | 0.0765 | 1.229 | 0.0502 |
| 2453796.3104 | 0.0872 | 0.944 | 0.0385 | 2453796.3095 | 0.0817 | 1.199 | 0.0490 |
| 2453796.3122 | 0.0925 | 0.923 | 0.0377 | 2453796.3113 | 0.0868 | 1.177 | 0.0481 |
| 2453796.3139 | 0.0976 | 0.897 | 0.0367 | 2453796.3130 | 0.0920 | 1.166 | 0.0476 |
| 2453796.3157 | 0.1028 | 0.889 | 0.0364 | 2453796.3148 | 0.0972 | 1.147 | 0.0468 |
| 2453796.3175 | 0.1080 | 0.873 | 0.0356 | 2453796.3166 | 0.1024 | 1.131 | 0.0462 |
| 2453796.3192 | 0.1132 | 0.867 | 0.0354 | 2453796.3183 | 0.1076 | 1.116 | 0.0456 |
| 2453796.3210 | 0.1184 | 0.844 | 0.0345 | 2453796.3201 | 0.1128 | 1.103 | 0.0450 |
| 2453796.3228 | 0.1235 | 0.832 | 0.0340 | 2453796.3219 | 0.1180 | 1.089 | 0.0445 |
| 2453796.3245 | 0.1287 | 0.818 | 0.0334 | 2453796.3237 | 0.1231 | 1.080 | 0.0441 |
| 2453796.3263 | 0.1339 | 0.808 | 0.0330 | 2453796.3254 | 0.1283 | 1.074 | 0.0439 |
| 2453796.3284 | 0.1399 | 0.795 | 0.0325 | 2453796.3275 | 0.1343 | 1.058 | 0.0432 |
| 2453796.3301 | 0.1451 | 0.788 | 0.0322 | 2453796.3292 | 0.1395 | 1.041 | 0.0425 |
| 2453796.3319 | 0.1503 | 0.776 | 0.0317 | 2453796.3310 | 0.1447 | 1.042 | 0.0425 |
| 2453796.3337 | 0.1555 | 0.765 | 0.0312 | 2453796.3328 | 0.1499 | 1.032 | 0.0421 |
| 2453796.3354 | 0.1607 | 0.757 | 0.0309 | 2453796.3346 | 0.1551 | 1.019 | 0.0416 |
| 2453796.3372 | 0.1658 | 0.751 | 0.0307 | 2453796.3381 | 0.1654 | 1000. | 0.0408 |
| 2453796.3390 | 0.1710 | 0.736 | 0.0301 | 2453796.3399 | 0.1706 | 0.993 | 0.0405 |
| 2453796.3407 | 0.1762 | 0.737 | 0.0301 | 2453796.3416 | 0.1757 | 0.984 | 0.0402 |
| 2453796.3425 | 0.1813 | 0.717 | 0.0293 | 2453796.3434 | 0.1809 | 0.980 | 0.0400 |
| 2453796.3443 | 0.1865 | 0.711 | 0.0290 | 2453796.3504 | 0.2016 | 0.953 | 0.0389 |
| 2453796.3460 | 0.1916 | 0.696 | 0.0284 | 2453796.3522 | 0.2068 | 0.954 | 0.0390 |
| 2453796.3478 | 0.1968 | 0.694 | 0.0283 | 2453796.3557 | 0.2171 | 0.956 | 0.0390 |
| 2453796.3496 | 0.2020 | 0.687 | 0.0281 | 2453796.3593 | 0.2275 | 0.957 | 0.0391 |
| 2453796.3513 | 0.2072 | 0.700 | 0.0286 | 2453796.3610 | 0.2326 | 0.965 | 0.0394 |
| 2453796.3531 | 0.2124 | 0.685 | 0.0280 | 2453796.3628 | 0.2378 | 0.967 | 0.0395 |
| 2453796.3549 | 0.2245 | 0.692 | 0.0283 | 2453796.3646 | 0.2431 | 0.967 | 0.0395 |
| 2453796.3566 | 0.2298 | 0.687 | 0.0281 | 2453796.3663 | 0.2482 | 0.962 | 0.0393 |
| 2453796.3584 | 0.2349 | 0.693 | 0.0283 | 2453796.3681 | 0.2533 | 0.965 | 0.0394 |
| 2453796.3602 | 0.2401 | 0.692 | 0.0283 | 2453796.3699 | 0.2586 | 0.965 | 0.0394 |
| 2453796.3619 | 0.2453 | 0.696 | 0.0284 | 2453796.3734 | 0.2689 | 0.973 | 0.0397 |
| 2453796.3637 | 0.2504 | 0.686 | 0.0280 | 2453796.3752 | 0.2740 | 0.976 | 0.0399 |
| 2453796.3655 | 0.2556 | 0.693 | 0.0283 | 2453796.3769 | 0.2793 | 0.980 | 0.0400 |
| 2453796.3672 | 0.2608 | 0.697 | 0.0285 | 2453796.3787 | 0.2844 | 0.991 | 0.0405 |
| 2453796.3690 | 0.2659 | 0.699 | 0.0285 | 2453796.3845 | 0.3013 | 0.984 | 0.0402 |
| 2453796.3708 | 0.2711 | 0.692 | 0.0283 | 2453796.3863 | 0.3065 | 0.989 | 0.0404 |
| 2453796.3725 | 0.2763 | 0.700 | 0.0286 | 2453796.3880 | 0.3117 | 0.994 | 0.0406 |
| 2453796.3743 | 0.2815 | 0.699 | 0.0285 | 2453796.3916 | 0.3221 | 1.006 | 0.0411 |
| 2453796.3760 | 0.2866 | 0.700 | 0.0286 | 2453796.3933 | 0.3272 | 1.020 | 0.0416 |
| 2453796.3778 | 0.2918 | 0.711 | 0.0290 | 2453796.3950 | 0.3323 | 1.028 | 0.0420 |
| 2453796.3796 | 0.2970 | 0.711 | 0.0290 | 2453796.3968 | 0.3375 | 1.035 | 0.0423 |
| 2453796.3818 | 0.3036 | 0.717 | 0.0293 | 2453796.3986 | 0.3426 | 1.046 | 0.0427 |
| 2453796.3836 | 0.3088 | 0.722 | 0.0295 | 2453796.4003 | 0.3478 | 1.048 | 0.0428 |
| 2453796.3854 | 0.3140 | 0.743 | 0.0303 | 2453796.4022 | 0.3531 | 1.052 | 0.0430 |
| 2453796.3871 | 0.3191 | 0.748 | 0.0305 | 2453796.4039 | 0.3583 | 1.069 | 0.0436 |
| 2453796.3889 | 0.3243 | 0.750 | 0.0306 | 2453796.4057 | 0.3635 | 1.078 | 0.0440 |
| 2453796.3907 | 0.3295 | 0.759 | 0.0310 | 2453796.4075 | 0.3686 | 1.096 | 0.0447 |
| 2453796.3924 | 0.3347 | 0.754 | 0.0308 | 2453796.4093 | 0.3739 | 1.107 | 0.0452 |
| 2453796.3942 | 0.3397 | 0.759 | 0.0310 | 2453796.4110 | 0.3790 | 1.121 | 0.0458 |
| 2453796.3959 | 0.3448 | 0.779 | 0.0318 | 2453803.3499 | 0.5870 | 1.224 | 0.0500 |
| 2453796.3977 | 0.3501 | 0.777 | 0.0317 | 2453803.3506 | 0.5891 | 1.211 | 0.0494 |
| 2453796.3995 | 0.3552 | 0.784 | 0.0320 | 2453803.3514 | 0.5915 | 1.216 | 0.0496 |
| 2453796.4013 | 0.3605 | 0.792 | 0.0323 | 2453803.3539 | 0.5987 | 1.175 | 0.0480 |
| 2453796.4030 | 0.3657 | 0.809 | 0.0330 | 2453803.3546 | 0.6007 | 1.173 | 0.0479 |
| 2453796.4048 | 0.3709 | 0.823 | 0.0336 | 2453803.3553 | 0.6028 | 1.186 | 0.0484 |
| 2453796.4066 | 0.3760 | 0.824 | 0.0336 | 2453803.3570 | 0.6078 | 1.175 | 0.0480 |
| 2453796.4084 | 0.3813 | 0.843 | 0.0344 | 2453803.3577 | 0.6098 | 1.147 | 0.0468 |
| 2453796.4101 | 0.3865 | 0.857 | 0.0350 | 2453803.3586 | 0.6125 | 1.122 | 0.0458 |
| 2453802.3605 | 0.6884 | 0.739 | 0.0302 | 2453803.3638 | 0.6278 | 1.061 | 0.0433 |
| 2453802.3616 | 0.6916 | 0.731 | 0.0298 | 2453803.3646 | 0.6299 | 1.117 | 0.0456 |
| 2453802.3626 | 0.6947 | 0.724 | 0.0296 | 2453803.3742 | 0.6582 | 1.006 | 0.0411 |
| 2453802.3637 | 0.6979 | 0.722 | 0.0295 | 2453803.3749 | 0.6602 | 1.052 | 0.0430 |
| 2453802.3688 | 0.7129 | 0.709 | 0.0290 | 2453803.3766 | 0.6651 | 1.041 | 0.0425 |
| 2453802.3699 | 0.7160 | 0.722 | 0.0296 | 2453803.3781 | 0.6696 | 1.013 | 0.0414 |
| 2453802.3710 | 0.7193 | 0.704 | 0.0287 | 2453803.3825 | 0.6826 | 0.993 | 0.0405 |
| 2453802.3721 | 0.7225 | 0.713 | 0.0291 | 2453803.3841 | 0.6873 | 0.993 | 0.0405 |
| 2453802.3732 | 0.7257 | 0.715 | 0.0292 | 72453803.393 | 0.7152 | 0.976 | 0.0399 |
| 2453802.3747 | 0.7299 | 0.707 | 0.0289 | 42453803.394 | 0.7173 | 0.957 | 0.0391 |
| 2453802.3760 | 0.7340 | 0.696 | 0.0284 | 2453803.4019 | 0.7393 | 0.962 | 0.0393 |
| 2453802.3830 | 0.7544 | 0.699 | 0.0285 | 32453803.410 | 0.7638 | 0.955 | 0.0390 |
| 2453802.3841 | 0.7576 | 0.703 | 0.0287 | 72453803.413 | 0.7738 | 0.965 | 0.0394 |
| 2453802.3896 | 0.7737 | 0.704 | 0.0287 | 2453803.4183 | 0.7875 | 0.968 | 0.0395 |
| 2453802.3907 | 0.7769 | 0.711 | 0.0290 | 702453803.42 | 0.8128 | 0.987 | 0.0403 |
| 2453802.3971 | 0.7957 | 0.717 | 0.0293 | 52453803.429 | 0.8202 | 0.977 | 0.0399 |
| 2453802.4065 | 0.8232 | 0.760 | 0.0310 | 2453803.4302 | 0.8223 | 0.991 | 0.0405 |
| 2453802.4098 | 0.8328 | 0.759 | 0.0310 | 72453803.436 | 0.8412 | 1.031 | 0.0421 |
| 2453802.4123 | 0.8403 | 0.772 | 0.0315 | 2453803.4380 | 0.8452 | 1.041 | 0.0425 |
| 2453802.4152 | 0.8487 | 0.786 | 0.0321 | 82453803.438 | 0.8473 | 1.067 | 0.0436 |
| 2453802.4186 | 0.8586 | 0.806 | 0.0329 | 52453803.439 | 0.8493 | 1.041 | 0.0425 |
| 2453802.4204 | 0.8638 | 0.815 | 0.0333 | 22453803.440 | 0.8514 | 1.044 | 0.0426 |
| 2453802.4215 | 0.8670 | 0.816 | 0.0333 | 62453803.447 | 0.8731 | 1.057 | 0.0432 |
| 2453802.4248 | 0.8767 | 0.829 | 0.0338 | 2453803.4491 | 0.8777 | 1.079 | 0.0441 |
| 2453802.4259 | 0.8799 | 0.851 | 0.0347 | 2453803.4565 | 0.8994 | 1.162 | 0.0474 |
| 2453802.4269 | 0.8831 | 0.845 | 0.0345 | 42453815.357 | 0.8837 | 1.109 | 0.0453 |
| 2453802.4285 | 0.8877 | 0.864 | 0.0353 | 72453815.359 | 0.8903 | 1.131 | 0.0462 |
| 2453803.3497 | 0.5863 | 0.947 | 0.0387 | 2453815.3619 | 0.8970 | 1.149 | 0.0469 |
| 2453803.3504 | 0.5885 | 0.938 | 0.0383 | 82453815.363 | 0.9023 | 1.163 | 0.0475 |
| 2453803.3512 | 0.5908 | 0.955 | 0.0390 | 2453815.3642 | 0.9036 | 1.190 | 0.0486 |
| 2453803.3537 | 0.5980 | 0.922 | 0.0376 | 72453815.364 | 0.9050 | 1.178 | 0.0481 |
| 2453803.3544 | 0.6000 | 0.897 | 0.0366 | 32453815.368 | 0.9156 | 1.217 | 0.0497 |
| 2453803.3551 | 0.6021 | 0.907 | 0.0370 | 82453815.368 | 0.9170 | 1.227 | 0.0501 |
| 2453803.3558 | 0.6041 | 0.910 | 0.0372 | 2453815.3692 | 0.9183 | 1.229 | 0.0502 |
| 2453803.3574 | 0.6091 | 0.887 | 0.0362 | 62453815.370 | 0.9223 | 1.252 | 0.0511 |
| 2453803.3622 | 0.6230 | 0.860 | 0.0351 | 52453815.371 | 0.9249 | 1.267 | 0.0517 |
| 2453803.3629 | 0.6251 | 0.850 | 0.0347 | 32453815.373 | 0.9303 | 1.275 | 0.0521 |
| 2453803.3636 | 0.6271 | 0.840 | 0.0343 | 82453815.373 | 0.9316 | 1.289 | 0.0526 |
| 2453803.3643 | 0.6292 | 0.814 | 0.0332 | 72453815.374 | 0.9343 | 1.311 | 0.0535 |
| 2453803.3696 | 0.6447 | 0.816 | 0.0333 | 2453815.3751 | 0.9356 | 1.316 | 0.0537 |
| 2453803.3704 | 0.6470 | 0.794 | 0.0324 | 2453815.3792 | 0.9476 | 1.399 | 0.0571 |
| 2453803.3714 | 0.6499 | 0.784 | 0.0320 | 72453815.384 | 0.9635 | 1.519 | 0.0620 |
| 2453803.3723 | 0.6525 | 0.811 | 0.0331 | 62453815.385 | 0.9662 | 1.533 | 0.0626 |
| 2453803.3730 | 0.6546 | 0.778 | 0.0318 | 2453815.3860 | 0.9676 | 1.513 | 0.0618 |
| 2453803.3739 | 0.6574 | 0.788 | 0.0322 | 52453815.386 | 0.9689 | 1.534 | 0.0626 |
| 2453803.3747 | 0.6595 | 0.786 | 0.0321 | 2453815.3869 | 0.9702 | 1.549 | 0.0632 |
| 2453803.3798 | 0.6745 | 0.751 | 0.0307 | 72453815.389 | 0.9782 | 1.566 | 0.0639 |
| 2453803.3805 | 0.6766 | 0.767 | 0.0313 | 2453815.3901 | 0.9795 | 1.580 | 0.0645 |
| 2453803.4300 | 0.8216 | 0.733 | 0.0299 | 2453815.3919 | 0.9848 | 1.592 | 0.0650 |
| 2453803.4489 | 0.8769 | 0.848 | 0.0346 | 2453815.3942 | 0.9915 | 1.616 | 0.0660 |
| 2453803.4504 | 0.8815 | 0.824 | 0.0336 | 72453815.394 | 0.9928 | 1.603 | 0.0654 |
| 2453803.4515 | 0.8847 | 0.797 | 0.0325 | 2453815.3951 | 0.9942 | 1.627 | 0.0664 |
| 2453803.4549 | 0.8945 | 0.911 | 0.0372 | 62453815.395 | 0.9955 | 1.624 | 0.0663 |
| 2453803.4556 | 0.8967 | 0.907 | 0.0370 | 2453815.3960 | 0.9968 | 1.623 | 0.0663 |
| 2453814.3484 | 0.9179 | 0.974 | 0.0398 | 52453815.396 | 0.9981 | 1.622 | 0.0662 |
| 2453814.3518 | 0.9276 | 1.025 | 0.0419 | 2453815.3969 | 0.9995 | 1.624 | 0.0663 |
| 2453814.3551 | 0.9374 | 1.075 | 0.0439 | 42453815.397 | 0.0009 | 1.611 | 0.0658 |
| 2453814.3585 | 0.9473 | 1.135 | 0.0463 | 2453815.4037 | 0.0194 | 1.592 | 0.0650 |
| 2453814.3618 | 0.9571 | 1.182 | 0.0483 | 2453815.4042 | 0.0208 | 1.577 | 0.0644 |
| 2453814.3652 | 0.9669 | 1.252 | 0.0511 | 72453815.404 | 0.0221 | 1.563 | 0.0638 |
| 2453814.3685 | 0.9767 | 1.319 | 0.0539 | 2453815.4069 | 0.0288 | 1.542 | 0.0630 |
| 2453814.3719 | 0.9865 | 1.349 | 0.0551 | 42453815.407 | 0.0301 | 1.525 | 0.0623 |
| 2453814.3752 | 0.9963 | 1.376 | 0.0562 | 2453815.4078 | 0.0314 | 1.536 | 0.0627 |
| 2453814.3785 | 0.0061 | 1.367 | 0.0558 | 32453815.408 | 0.0328 | 1.523 | 0.0622 |
| 2453814.3819 | 0.0159 | 1.354 | 0.0553 | 82453815.408 | 0.0341 | 1.521 | 0.0621 |
| 2453814.3852 | 0.0257 | 1.323 | 0.0540 | 2453815.4092 | 0.0354 | 1.472 | 0.0601 |
| 2453814.3886 | 0.0355 | 1.278 | 0.0522 | 72453815.409 | 0.0368 | 1.500 | 0.0612 |
| 2453814.3919 | 0.0453 | 1.196 | 0.0488 | 62453815.410 | 0.0394 | 1.490 | 0.0608 |
| 2453814.3953 | 0.0551 | 1.125 | 0.0459 | 2453815.4110 | 0.0408 | 1.442 | 0.0589 |
| 2453814.3986 | 0.0649 | 1.078 | 0.0440 | 52453815.411 | 0.0421 | 1.458 | 0.0595 |
| 2453814.4020 | 0.0747 | 1.002 | 0.0409 | 2453815.4119 | 0.0434 | 1.458 | 0.0595 |
| 2453814.4053 | 0.0845 | 0.987 | 0.0403 | 2453815.4128 | 0.0461 | 1.419 | 0.0579 |
| 2453814.4087 | 0.0943 | 0.947 | 0.0387 | 32453815.413 | 0.0474 | 1.416 | 0.0578 |
| 2453814.4120 | 0.1041 | 0.913 | 0.0373 | 82453815.413 | 0.0488 | 1.405 | 0.0574 |
| 2453814.4154 | 0.1139 | 0.867 | 0.0354 | 2453815.4142 | 0.0501 | 1.412 | 0.0577 |
| 2453814.4187 | 0.1237 | 0.843 | 0.0344 | 72453815.414 | 0.0515 | 1.381 | 0.0564 |
| 2453814.4221 | 0.1336 | 0.830 | 0.0339 | 2453815.4151 | 0.0528 | 1.387 | 0.0566 |
| 2453814.4254 | 0.1434 | 0.777 | 0.0317 | 62453815.415 | 0.0541 | 1.363 | 0.0556 |
| 2453814.4288 | 0.1532 | 0.781 | 0.0319 | 2453815.4160 | 0.0554 | 1.357 | 0.0554 |
| 2453835.3175 | 0.3680 | 0.812 | 0.0332 | 52453815.416 | 0.0567 | 1.373 | 0.0561 |
| 2453835.3195 | 0.3740 | 0.797 | 0.0325 | 2453815.4169 | 0.0581 | 1.315 | 0.0537 |
| 2453835.3215 | 0.3797 | 0.818 | 0.0334 | 2453815.4210 | 0.0701 | 1.273 | 0.0520 |
| 2453835.3234 | 0.3854 | 0.812 | 0.0332 | 52453815.421 | 0.0714 | 1.277 | 0.0521 |
| 2453835.3254 | 0.3912 | 0.861 | 0.0352 | 2453835.3307 | 0.3967 | 1.160 | 0.0474 |
| 2453835.3274 | 0.3969 | 0.884 | 0.0361 | 72453835.332 | 0.4025 | 1.177 | 0.0481 |
| 2453835.3293 | 0.4026 | 0.861 | 0.0352 | 2453835.3346 | 0.4082 | 1.184 | 0.0483 |
| 2453835.3312 | 0.4083 | 0.889 | 0.0363 | 62453835.336 | 0.4140 | 1.213 | 0.0495 |
| 2453835.3332 | 0.4141 | 0.895 | 0.0365 | 2453835.3385 | 0.4197 | 1.253 | 0.0512 |
| 2453835.3352 | 0.4198 | 0.926 | 0.0378 | 52453835.340 | 0.4254 | 1.253 | 0.0512 |
| 2453835.3371 | 0.4255 | 0.975 | 0.0398 | 52453835.342 | 0.4312 | 1.302 | 0.0532 |
| 2453835.3391 | 0.4312 | 0.978 | 0.0399 | 2453835.3444 | 0.4369 | 1.325 | 0.0541 |
| 2453835.3410 | 0.4370 | 1.038 | 0.0424 | 42453835.346 | 0.4427 | 1.342 | 0.0548 |
| 2453835.3430 | 0.4428 | 1.082 | 0.0442 | 42453835.348 | 0.4484 | 1.423 | 0.0581 |
| 2453835.3450 | 0.4485 | 1.108 | 0.0452 | 32453835.350 | 0.4541 | 1.445 | 0.0590 |
| 2453835.3469 | 0.4542 | 1.113 | 0.0454 | 32453835.352 | 0.4598 | 1.480 | 0.0604 |
| 2453835.3489 | 0.4600 | 1.112 | 0.0454 | 2453835.3542 | 0.4656 | 1.494 | 0.0610 |
| 2453835.3508 | 0.4657 | 1.207 | 0.0493 | 22453835.356 | 0.4713 | 1.521 | 0.0621 |
| 2453835.3528 | 0.4714 | 1.233 | 0.0503 | 2453835.3581 | 0.4770 | 1.578 | 0.0644 |
| 2453835.3548 | 0.4772 | 1.209 | 0.0494 | 12453835.360 | 0.4828 | 1.581 | 0.0645 |
| 2453835.3567 | 0.4829 | 1.292 | 0.0528 | 2453835.3659 | 0.4999 | 1.583 | 0.0646 |
| 2453835.3626 | 0.5001 | 1.317 | 0.0538 | 92453835.367 | 0.5057 | 1.573 | 0.0642 |
| 2453835.3665 | 0.5115 | 1.300 | 0.0531 | 92453835.369 | 0.5114 | 1.570 | 0.0641 |

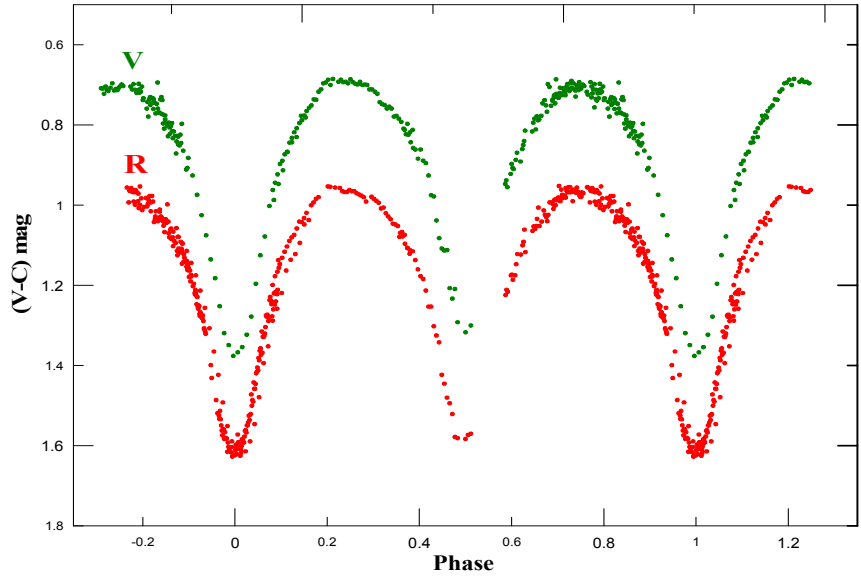

Figure 1. Light curves of EQ Tau in V and R filters.

Table 3. Light minima of EQ Tau.

| HJD | Error | Min | Filter |
|---|---|---|---|
| 2453814.3356 | ±0.0002 | I | V |
| 2453814.3358 | ±0.0005 | I | B |
| 2453815.3551 | ±0.0003 | I | R |

Table 4. Times of minimum light for EQ Tau.

| JD(Hel) | E | (O-C) | (O-C)p | Min | Method | Ref. | JD(Hel) | E | (O-C) | (O-C)p | Min | Method | Ref. |
|---|---|---|---|---|---|---|---|---|---|---|---|---|---|
| 2451849.1874 | 34117 | -0.0148 | -0.0012 | p | ccd | 1 | 2455500.2506 | 44813 | -0.0093 | 0.0012 | P | ccd | 35 |
| 2451849.1875 | 34117 | -0.0147 | -0.0011 | p | ccd | 1 | 2455500.4222 | 44813.5 | -0.0083 | 0.0021 | S | ccd | 35 |
| 2452219.5485 | 35202 | -0.0162 | -0.0020 | p | ccd | 2 | 2455500.4223 | 44813.5 | -0.0082 | 0.0022 | S | ccd | 35 |
| 2452219.5486 | 35202 | -0.0161 | -0.0019 | p | ccd | 2 | 2455500.4223 | 44813.5 | -0.0082 | 0.0022 | S | ccd | 35 |
| 2452219.5490 | 35202 | -0.0157 | -0.0015 | p | ccd | 2 | 2455511.1719 | 44845 | -0.0111 | -0.0007 | S | ccd | 35 |
| 2452296.6934 | 35428 | -0.0160 | -0.0017 | p | ccd | 3 | 2455511.6870 | 44846.5 | -0.0080 | 0.0015 | S | ccd | 14 |
| 2452614.6597 | 36359.5 | -0.0153 | -0.0007 | S | ccd | 3 | 2455520.7310 | 44873 | -0.0097 | -0.0003 | p | ccd | 15 |
| 2452620.6330 | 36377 | -0.0156 | -0.0009 | p | ccd | 3 | 2455543.0906 | 44938.5 | -0.0084 | 0.0019 | S | ccd | 35 |
| 2452684.6360 | 36564.5 | -0.0153 | -0.0006 | S | ccd | 3 | 2455539.6771 | 44928.5 | -0.0084 | 0.0009 | S | ccd | 16 |
| 2453019.1580 | 37544.5 | -0.0143 | 0.0006 | S | ccd | 1 | 2455551.1116 | 44962 | -0.0091 | 0.0002 | p | ccd | 17 |
| 2453351.9778 | 38519.5 | -0.0088 | 0.0061 | S | ccd | 1 | 2455554.0125 | 44970.5 | -0.0097 | -0.0004 | S | ccd | 17 |
| 2453352.1414 | 38520 | -0.0158 | -0.0010 | p | ccd | 1 | 2455554.1835 | 44971 | -0.0093 | -00010 | p | ccd | 17 |
| 2453352.1416 | 38520 | -0.0156 | -0.0008 | p | ccd | 1 | 2455570.5690 | 45019 | -0.0085 | 0.0006 | p | vis | 18 |
| 2453358.1162 | 38537.5 | -0.0146 | 0.0002 | S | ccd | 1 | 2455643.2755 | 45232 | -0.0091 | -0.0004 | p | ccd | 19 |
| 2453358.1164 | 38537.5 | -0.0144 | 0.0004 | S | ccd | 1 | 2455643.2757 | 45232 | -0.0089 | -0.0002 | p | ccd | 19 |
| 2453814.3773 | 39874 | 0.0350 | 0.0495 | p | ccd | 4 | 2455643.2758 | 45232 | -0.0088 | -0.0001 | p | ccd | 19 |
| 2453814.3774 | 39874 | 0.0351 | 0.0496 | p | ccd | 4 | 2455844.8368 | 45822.5 | -0.0138 | -0.0062 | S | ccd | 20 |
| 2453815.3957 | 39877 | 0.0293 | 0.0438 | p | ccd | 4 | 2455846.8886 | 45828.5 | -0.0101 | -0.0025 | S | ccd | 21 |
| 2453835.3610 | 39935.5 | 0.0258 | 0.0402 | S | ccd | 4 | 2455865.6584 | 45883.5 | -0.0144 | -0.0069 | S | ccd | 22 |
| 2453835.3628 | 39936 | -0.1431 | -0.1286 | p | ccd | 4 | 2455885.6270 | 45942 | -0.0147 | -0.0073 | p | vis | 23 |
| 2454057.3698 | 40586 | -0.0123 | 0.0019 | p | ccd | 1 | 2455901.6750 | 45989 | -0.0100 | -0.0027 | p | ccd | 24 |
| 2454389.5007 | 41559 | -0.0129 | 0.0006 | p | ccd | 5 | 2455910.0387 | 46013.5 | -0.0094 | -0.0021 | S | ccd | 25 |
| 2454389.5007 | 41559 | -0.0129 | 0.0006 | p | ccd | 5 | 2455910.2091 | 46014 | -0.0096 | -0.0024 | p | ccd | 25 |

| | | | | | | | | | | | | |
|---|---|---|---|---|---|---|---|---|---|---|---|---|
| 2454389.5009 | 41559 | -0.0127 | 0.0008 | p | ccd | 5 | 2455921.6398 | 46047.5 | -0.0141 | -0.0069 | S | ccd | 26 |
| 2454506.2428 | 41901 | -0.0118 | 0.0014 | p | ccd | 5 | 2455937.5132 | 46094 | -0.0134 | -0.0063 | p | ccd | 26 |
| 2454506.2435 | 41901 | -0.0111 | 0.0021 | p | ccd | 5 | 2455963.6229 | 46170.5 | -0.0168 | -0.0099 | S | ccd | 27 |
| 2454509.3146 | 41910 | -0.0122 | 0.0011 | p | ccd | 5 | 2456182.7685 | 46812.5 | -0.0166 | -0.0111 | S | ccd | 28 |
| 2454509.3147 | 41910 | -0.0121 | 0.0012 | p | ccd | 5 | 2456222.3719 | 46928.5 | -0.0095 | -0.0043 | S | ccd | 29 |
| 2454509.3147 | 41910 | -0.0121 | 0.0012 | p | ccd | 5 | 2456243.6999 | 46991 | -0.0158 | -0.0107 | p | ccd | 30 |
| 2455116.7407 | 43689.5 | -0.0147 | -0.0035 | S | ccd | 6 | 2456251.3860 | 47013.5 | -0.0100 | -0.0050 | S | ccd | 31 |
| 2455148.3213 | 43782 | -0.0088 | 0.0028 | p | ccd | 35 | 2456273.0610 | 47077 | -0.0106 | -0.0035 | p | ccd | 35 |
| 2455175.4577 | 43861.5 | -0.0096 | 0.0014 | S | ccd | 7 | 2456273.0612 | 47077 | -0.0104 | -0.0033 | p | ccd | 35 |
| 2455175.4584 | 43861.5 | -0.0089 | 0.0021 | S | ccd | 7 | 2456276.1329 | 47086 | -0.0108 | -0.0037 | p | ccd | 35 |
| 2455175.4594 | 43861.5 | -0.0079 | 0.0031 | S | ccd | 7 | 2456276.1333 | 47086 | -0.0104 | -0.0033 | p | ccd | 35 |
| 2455175.4594 | 43861.5 | -0.0079 | 0.0031 | S | ccd | 7 | 2456276.6458 | 47087.5 | -0.0099 | -0.0052 | S | ccd | 32 |
| 2455175.6270 | 43862 | -0.0110 | 0.0001 | p | ccd | 7 | 2456290.6423 | 47128.5 | -0.0087 | -0.0040 | S | ccd | 33 |
| 2455175.6286 | 43862 | -0.0094 | 0.0017 | p | ccd | 7 | 2456291.4953 | 47131 | -0.0091 | -0.0044 | p | ccd | 34 |
| 2455186.3807 | 43893.5 | -0.0097 | 0.0013 | S | ccd | 8 | 2456301.0512 | 47159 | -0.0109 | -0.0040 | p | ccd | 35 |
| 2455192.6950 | 43912 | -0.0104 | 0.0006 | p | ccd | 9 | 2456301.0514 | 47159 | -0.0107 | -0.0038 | p | ccd | 35 |
| 2455219.3202 | 43990 | -0.0103 | 0.0005 | p | ccd | 8 | 2456301.0514 | 47159 | -0.0107 | -0.0038 | p | ccd | 35 |
| 2455259.5991 | 44108 | -0.0105 | 0.0002 | p | ccd | 10 | 2456301.0516 | 47159 | -0.0105 | -0.0036 | p | ccd | 35 |
| 2455453.4850 | 44676 | -0.0102 | -0.0004 | p | ccd | 8 | 2456301.2223 | 47159.5 | -0.0105 | -0.0035 | S | ccd | 35 |
| 2455460.3146 | 44696 | -0.0075 | 0.0031 | p | ccd | 35 | 2456301.2223 | 47159.5 | -0.0105 | -0.0035 | S | ccd | 35 |
| 2455460.3147 | 44696 | -0.0074 | 0.0032 | p | ccd | 35 | 2456301.2224 | 47159.5 | -0.0104 | -0.0034 | S | ccd | 35 |
| 2455480.4519 | 44755 | -0.0098 | -0.0002 | p | ccd | 11 | 2456301.2225 | 47159.5 | -0.0103 | -0.0033 | S | ccd | 35 |
| 2455485.7443 | 44770.5 | -0.0083 | 0.0013 | s | ccd | 12 | 2456549.5516 | 47887 | -0.0118 | -0.0091 | p | ccd | 36 |
| 2455498.8846 | 44809 | -0.0099 | -0.0003 | p | ccd | 13 | 2456556.5502 | 47907.5 | -0.0109 | -0.0082 | S | ccd | 36 |
| 2455499.3979 | 44810.5 | -0.0086 | 0.0010 | S | ccd | 8 | 2456592.3899 | 48012.5 | -0.0127 | -0.0103 | S | ccd | 37 |
| 2455500.2505 | 44813 | -0.0090 | 0.0011 | p | ccd | 35 | 2456613.7242 | 48075 | -0.0127 | -0.0105 | p | ccd | 38 |
| 2455500.2505 | 44813 | -0.0094 | 0.0011 | p | ccd | 35 | 2456619.8641 | 48093 | -0.0170 | -0.0149 | p | ccd | 39 |

Table 5 Period behavior for the system EQ Tau

| Parameters | Intervals (2400000+) | | | |
|---|---|---|---|---|
| | $E_0$ to $E_1$ 30647 - 34389 | $E_1$ to $E_2$ 34389 – 43483 | $E_2$ to $E_3$ 43483 -52185 | $E_3$ to $E_4$ 52185 - 56301 |
| $\Delta E$ (day) | 3742 | 9094 | 8702 | 4116 |
| Period (day) | 0.341345116 | 0.341348869 | 0.341346785 | 0.341348580 |
| $\Delta P$ (day) | $-2.831 \times 10^{-6}$ | $9.220 \times 10^{-7}$ | $-1.162 \times 10^{-6}$ | $6.327 \times 10^{-7}$ |
| $\Delta P/P$ | $-8.293 \times 10^{-6}$ | $2.701 \times 10^{-6}$ | $-3.404 \times 10^{-6}$ | $1.854 \times 10^{-6}$ |
| $\Delta P/ \Delta E$ (d/cycle) | $-7.566 \times 10^{-10}$ | $1.014 \times 10^{-10}$ | $-1.335 \times 10^{-10}$ | $1.537 \times 10^{-10}$ |
| Epoch (2400000+) | 40203.3664 | 40203.4278 | 40203.4582 | 40203.3933 |
| SD | 0.00165 | 0.00269 | 0.00265 | 0.00116 |
| r | 0.99790 | 0.88708 | 0.94931 | 0.89122 |
| Residual sum of square | 0.000003 | 0.00015 | 0.00032 | 0.00026 |
| R_Squared | 0.99580 | 0.78691 | 0.90119 | 0.79427 |

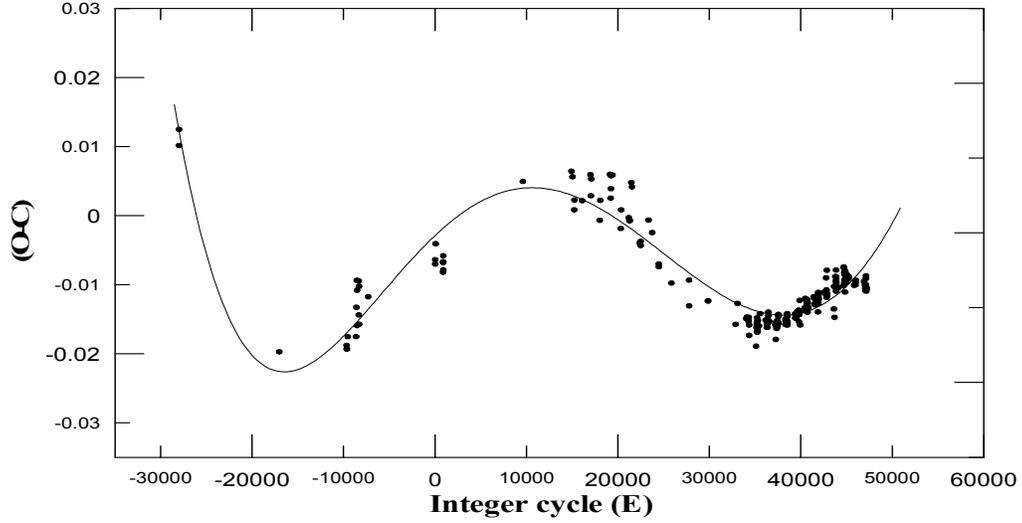

Figure 2. Period behavior of EQ Tau

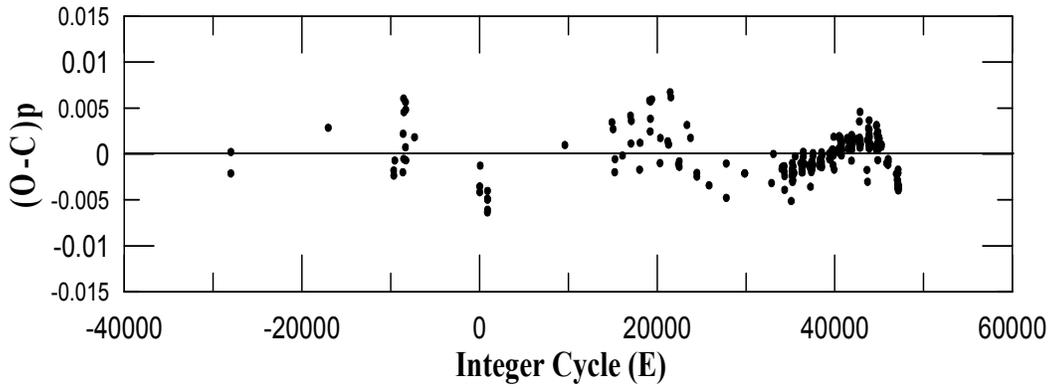

Figure 3. Calculated residuals from the polynomial ephemeris

Table 6. Light curve parameters for EQ Tau

| JD | Date | $D_{max}$ (mag) | $D_{min}$ (mag) | $A_p$ (mag) | $A_s$ (mag) | Ref. |
|---|---|---|---|---|---|---|
| 2451849 | 2000.8 | 0.0040±0.0002 | 0.0640±0.0026 | 0.6860±0.0280 | 0.6220±0.0254 | 1 |
| 2451941 | 2001.08 | -0.0290±0.0012 | 0.0840±0.0034 | 0.7160±0.0292 | 0.6320±0.0258 | 1 |
| 2452072 | 2001.5 | 0.0170±0.0007 | 0.0740±0.0030 | 0.6780±0.0277 | 0.6040±0.0247 | 2 |
| 2452240 | 2001.9 | -0.0300±0.0012 | 0.1100±0.0045 | 0.7700±0.0314 | 0.6600±0.0269 | 3 |
| 2452639 | 2003 | -0.0350±0.0014 | 0.1000±0.0041 | 0.7750±0.0316 | 0.6750±0.0276 | 4 |
| 2453355 | 2005 | -0.0490±0.0020 | 0.0690±0.0028 | 0.7070±0.0289 | 0.6380±0.0261 | 1 |
| 2453770 | 2006.1 | -0.0400±0.0016 | 0.0350±0.0014 | 0.4450±0.0182 | 0.4100±0.0167 | 5 |
| 2453821 | 2006.3 | 0.0030±0.0001 | 0.0545±0.0022 | 0.6855±0.0280 | 0.6310±0.0258 | 6 |
| 2454514 | 2008.1 | -0.0450±0.0018 | 0.0280±0.0011 | 0.4850±0.0198 | 0.5130±0.0209 | 7 |
| 2455500 | 2010.8 | -0.0250±0.0010 | 0.0400±0.0016 | 0.7400±0.0302 | 0.7000±0.0286 | 8 |
| 2456301 | 2013.1 | -0.0250±0.0010 | 0.0700±0.0029 | 0.7400±0.0302 | 0.6700±0.0274 | 8 |

**Reference:** Yuan & Qian (2007), Pribulla&Vanko (2002), Yang&Liu (2002), Hrivnak et al. (2006), Alton (2006), This Paper, Alton (2008), Li et al. (2014).

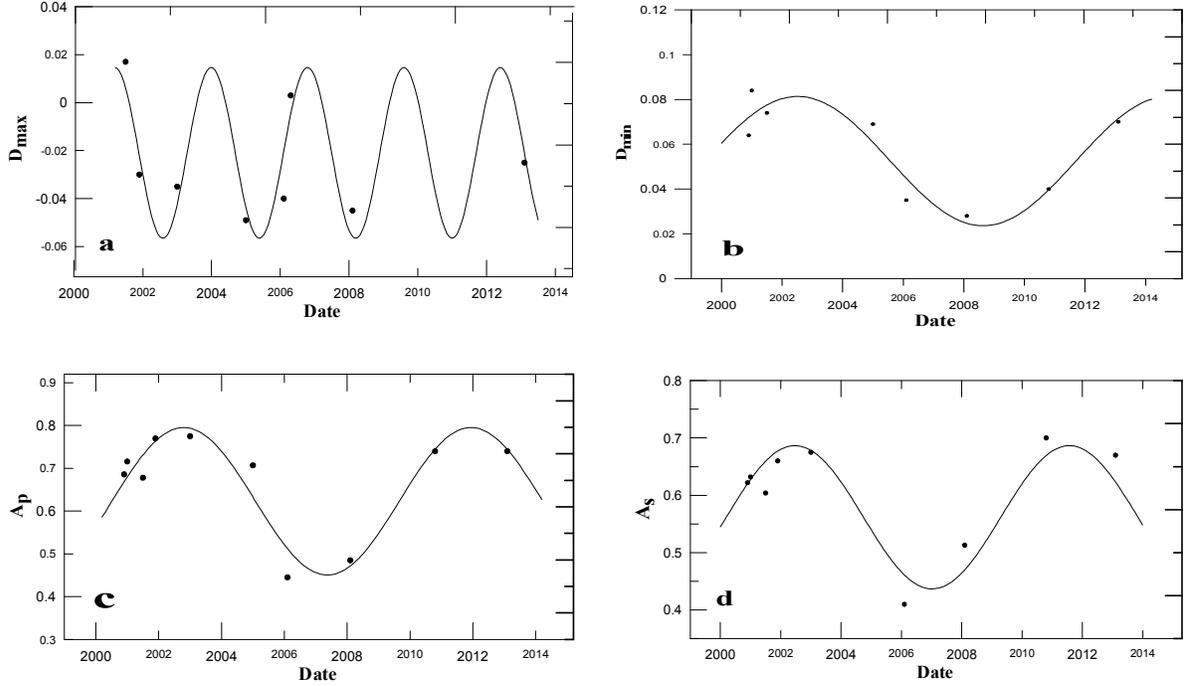

Figure 4 Variations of $D_{max}$, $D_{min}$, $A_p$, and $A_s$, for the system EQ Tau in V filter

Table 7. Photometric solution EQ Tau.

| Parameter | Element |
|---|---|
| $i$ ($^0$) | 82.04±0.85 |
| $g_1 = g_2$ | 0.5 |
| $A_1 = A_2$ | 0.32 |
| q ($M_2 / M_1$) | 0.4453±0.0014 |
| $\Omega_1 = \Omega_2$ | 2.7399±0.0047 |
| $\Omega_{in}$ | 2.7691 |
| $\Omega_{out}$ | 2.5001 |
| $T_1$ (K) | 5800  Fixed |
| $T_2$ (K) | 5701±5 |
| $r_1$ pole | 0.4291±0.0002 |
| $r_1$ side | 0.4580±0.0003 |
| $r_1$ back | 0.4874±0.0005 |
| $r_2$ pole | 0.2962±0.0006 |
| $r_2$ side | 0.3097±0.0007 |
| $r_2$ back | 0.3460±0.0013 |
| ***Spot parameters for star 1*** | |
| ***Spot A:*** | |
| Co-latitude (deg) | 135.3±5.524 |
| Longitude (deg) | 146.7±5.989 |
| Spot radius (deg) | 9.100±0.372 |
| Temp. factor | 1.100±0.045 |
| ***Spot B:*** | |
| Co-latitude (deg) | 122±4.981 |
| Longitude (deg) | 75±3.062 |
| Spot radius (deg) | 11.24±0.459 |
| Temp. factor | 1.35±0.055 |
| ***Spot parameters for star 2*** | |
| Co-latitude (deg) | 118±4.817 |
| Longitude (deg) | 97.7±3.989 |
| Spot radius (deg) | 15.3±0.625 |
| Temp. factor | 1.43±0.058 |
| $\sum (O-C)^2$ | 0.04936 |

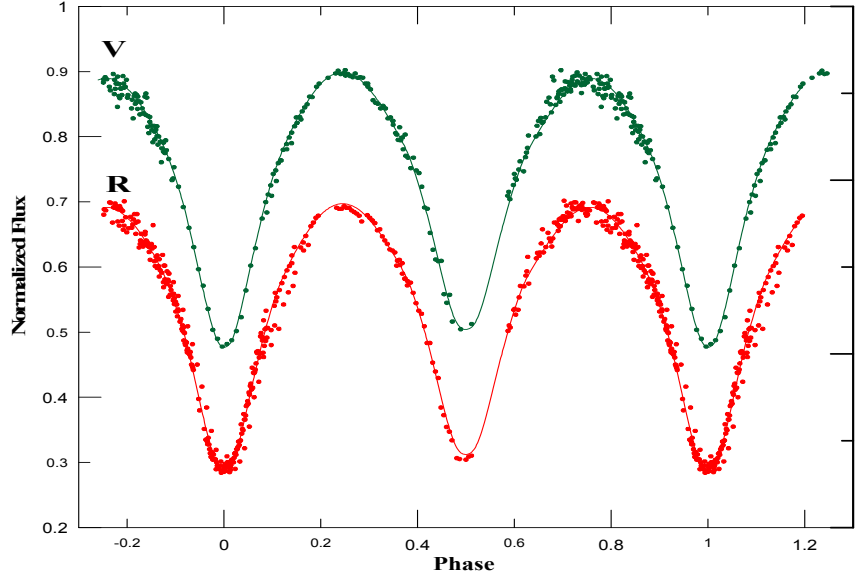

Figure 5, Observed light curves (filled circles), and the synthetic curves (solid line)

**Table (8): Physical parameters of EQ Tau.**

| Parameter | | | | | | | | Ref. |
|---|---|---|---|---|---|---|---|---|
| $M_1(M_\odot)$ | $M_2(M_\odot)$ | $R_1(R_\odot)$ | $R_2(R_\odot)$ | $L_1(L_\odot)$ | $L_2(L_\odot)$ | $\log T_1$ | $\log T_2$ | |
| 1.320 | 0.590 | 1.160 | 0.820 | 1.350 | 0.640 | 3.763 | 3.758 | 1 |
| 1.217 | 0.537 | 1.139 | 0.786 | 1.370 | 0.649 | 3.768 | 3.767 | 2 |
| 1.280 | 0.570 | 1.170 | 0.810 | 1.390 | 0.630 | 3.763 | 3.758 | 3 |
| 1.230 | 0.540 | 1.140 | 0.790 | 1.330 | 0.610 | 3.763 | 3.760 | 4 |
| 1.214 | 0.541 | 1.136 | 0.787 | 1.310 | 0.600 | 3.763 | 3.756 | 5 |

Reference: 1- Yang and Liu (2002), 2- Vanko et al (2004), 3- Hrivnak et al. (2006),
4- Yuang and Qian (2007), 5- This Paper

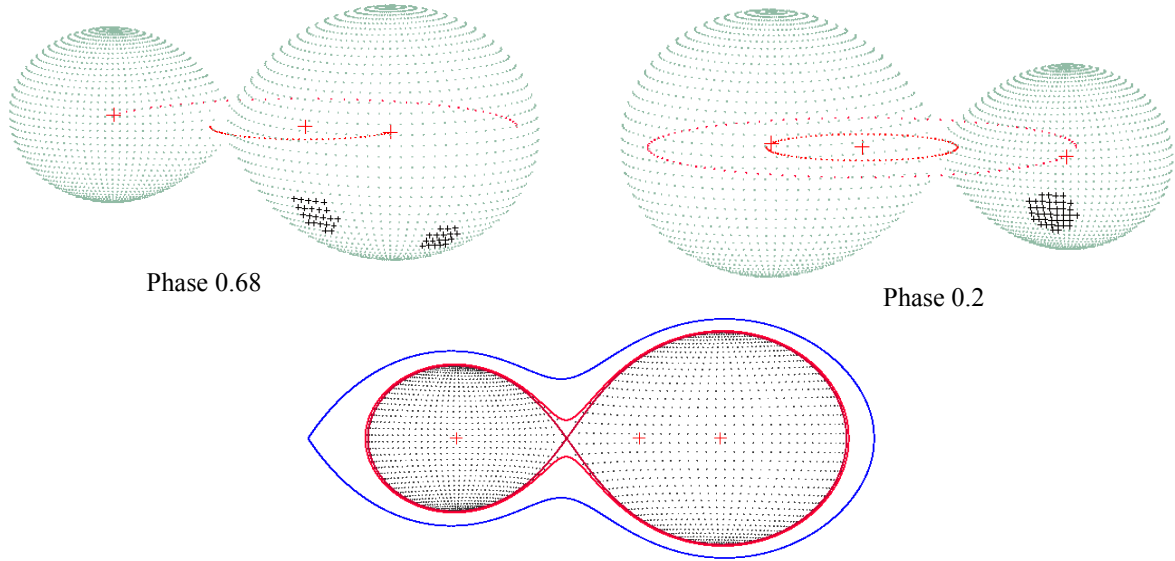

Figure 6 Geometric structure of the binary system EQ Tau.

Table (9): Physical parameters of the eight W-type contact binaries

| Star Name | Parameter | | | | | | | | Ref. |
|---|---|---|---|---|---|---|---|---|---|
| | $M_1(M_\odot)$ | $M_2(M_\odot)$ | $R_1(R_\odot)$ | $R_2(R_\odot)$ | $L_1(L_\odot)$ | $L_2(L_\odot)$ | $T_1(T_\odot)$ | $T_2(T_\odot)$ | |
| TY Boo | 1.03 | 0.48 | 1.02 | 0.72 | 0.5 | 0.89 | 0.99 | 0.95 | 1 |
| AW UMa | 1.6 | 0.121 | 1.786 | 0.739 | 7.47 | 0.804 | 1.242 | 1.218 | 2 |
| AD Cnc | 0.93 | 0.58 | 0.89 | 0.72 | 0.39 | 0.33 | 0.804 | 0.860 | 3 |
| TX Cnc | 1.37 | 0.82 | 1.23 | 0.96 | 2.2 | 1.38 | 1.056 | 1.066 | 3 |
| RZ Com | 1.03 | 0.45 | 1.06 | 0.71 | 0.93 | 0.43 | 0.916 | 0.927 | 3 |
| LS Del | 1.06 | 0.60 | 1.09 | 0.83 | 1.12 | 0.69 | 0.950 | 0.963 | 3 |
| BB Peg | 1.16 | 0.47 | 1.21 | 0.78 | 1.58 | 0.81 | 0.980 | 1.033 | 3 |
| AA UMa | 1.26 | 0.69 | 1.40 | 1.10 | 2.17 | 1.43 | 0.988 | 1.005 | 3 |

**Reference: 1- Elkhateeb et al. (2014), 2- Elkhateeb and Nouh (2014), 3- Macroni and van't Veer (1996).**

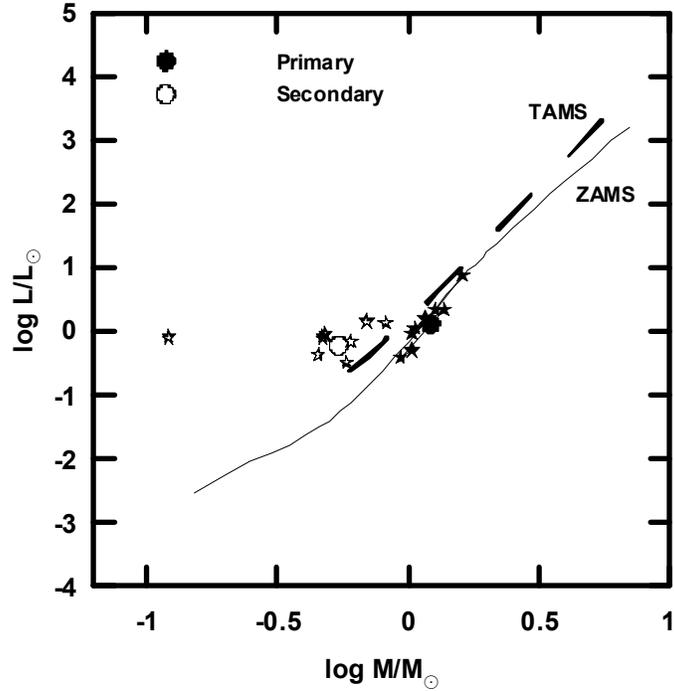

Figures (7): The position of the components of EQ Tau on the mass–luminosity diagram. The filled circle denotes the primary and the open circle represents the secondary. The other symbols denote the sample of the selected W-type systems listed in Table (9).

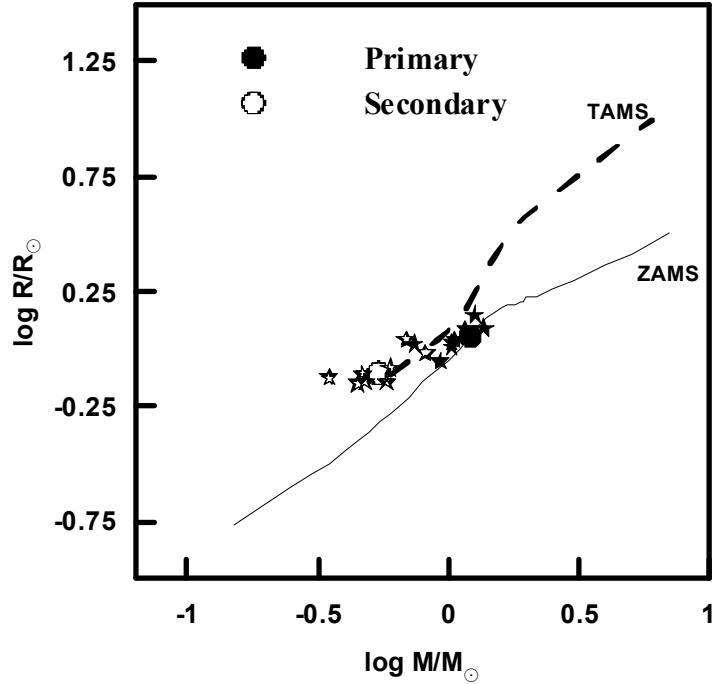

Figures (8): The position of the components of EQ Tau on the mass–radius diagram. The filled circle denotes the primary and the open circle represents the secondary. The other symbols denote the sample of the selected W-type systems listed in Table (9).

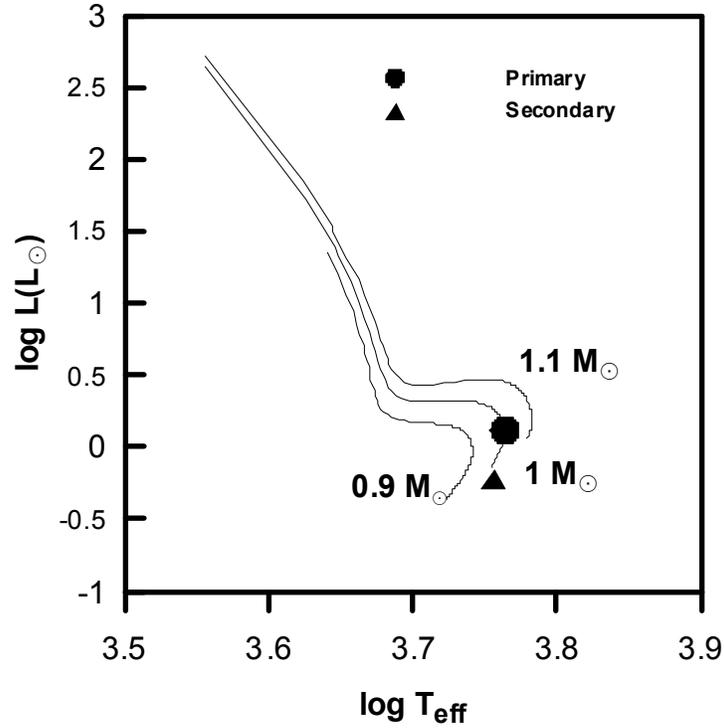

Figure (9). The position of the components of EQ Tau on the effective temperature – luminosity diagram of Estkron et al. (2012) .

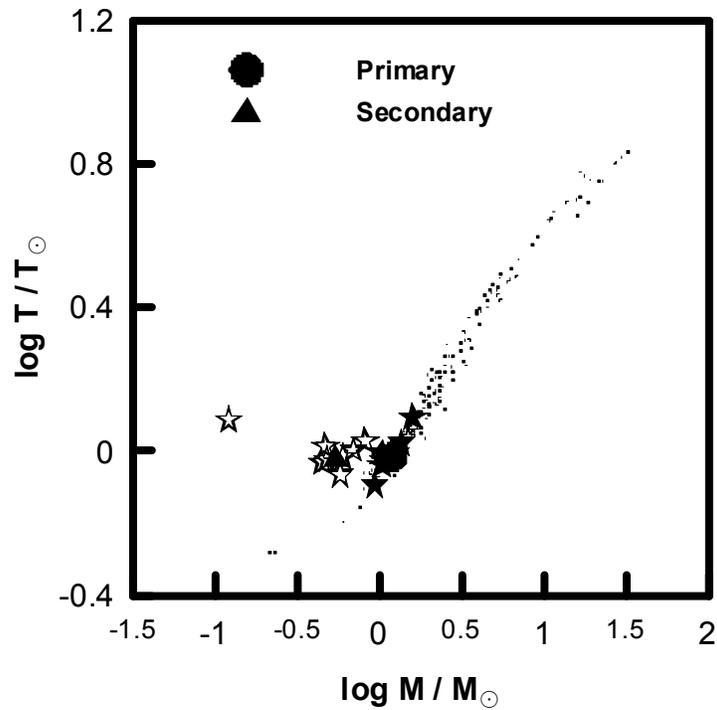

Figure (10). Position of the components of EQ Tau on the empirical mass-Teff relation for low-intermediate mass stars by Malkov (2007).